# A bandgap phenomenon in non-periodic plasmonic waveguides

Viacheslav Shaidiuk[‡], Sergey G. Menabde[‡] and Namkyoo Park*

*Photonic Systems Laboratory, School of EECS, Seoul National University, Seoul 151-744, Korea*

[‡] *Authors with equal contribution*

[*]*nkpark@snu.ac.kr*

**Abstract:** The phenomenon of a dispersion bandgap opening between low-loss spectral windows of odd and even plasmonic modes in a layered insulator-metal-insulator plasmonic waveguide is introduced. Beginning with a three layer plasmonic dispersion relation, we explain and numerically confirm the existence of the plasmonic bandgap, and investigate its properties at a very broad spectrum range from ultraviolet to far infrared. The nature of the observed bandgap opening is explained in terms of the near-zero value of an effective permittivity for plasmonic modes in the waveguide. The adjustment of the plasmonic bandgap spectrum is demonstrated with the structural modification of the plasmonic waveguide. As an application example, we illustrate a new concept of coupling control between surface plasmons and free-space excitation waves, by employing a tapered non-adiabatic insulator-metal-insulator waveguide.

## 1. Introduction

The interaction of light with matter has been a subject of thorough study for several decades. However, the idea of light manipulation using engineered materials (also called metamaterials) with artificially designed electromagnetic properties has only recently been realized [1-3]. While metamaterials has attracted a great deal of attention from optics and photonics communities, other types of structures also have exotic properties not conventionally observed in nature. For example, simple periodic media can exhibit a photonic and $\bar{n}$-zero bandgap [4,5], and extraordinary light transmission through a subwavelength size slit in a metallic sheet has been demonstrated [6]. Furthermore, within the larger frame of metal-optics, plasmonic structures allow light manipulation at subwavelength scale and offer equivalent exotic properties such as negative effective index in three layer (3L) plasmonic waveguides [7,8] or negative refraction in a stack of bilayer plasmonic waveguides [9-11]. These structures are relatively easy to analyze, design, and manufacture, and are a feasible and powerful tool for subwavelength light manipulation. Nevertheless, comprehensive study of their properties is yet to be fulfilled.

The elementary building block of the plasmonic structures mentioned above is the single insulator-metal (IM) interface supporting propagating plasmonic modes. The single IM interface plasmonic response is generally characterized by a mode with a relatively low effective index $k_p/k_0 < 10$ (where $k_0$ is an excitation wavevector in free space and $k_p$ is a plasmonic wavevector). At the same time, high effective index plasmonic modes of $k_p/k_0 > 10$, providing a pathway towards plasmon excitation on low dimensional materials (graphene, $MoS_2$, etc.) [11-13], are supported by 3L plasmonic waveguides with relatively thin metal or insulator film as the middle layer (i.e. insulator-metal-insulator (IMI) or metal-insulator-metal (MIM)), and can be excited via, for example, tapered waveguide structures [13,14], near-field optical excitation [15,16] or by surface sculpting and multiple incident beams [17-19]. However, while many studies have been conducted on 3L waveguide structures, the majority of works focus on symmetric waveguides and their plasmonic properties: mode effective index as a function of structural parameters, field distribution, penetration depth or propagation length [7,13,20-29] and missing in depth analysis on the plasmonic mode dispersion $k_p(k_0)$ at broader frequency range, especially in asymmetric 3L waveguides (i.e. when the middle layer is sandwiched between materials with different properties).

In this paper, we study plasmonic dispersion in *asymmetric* 3L waveguides and investigate newly observed phenomenon of bandgap opening in the domain of high effective index modes. First, we derive the dispersion of two-mode plasmonic response in 3L IMI asymmetric waveguides, which was previously shown only for the symmetric case [7, 8, 20, 23]. Starting from the obtained explicit dispersion relations for these two co-existing odd and even modes, we then analytically demonstrate the existence of a spectral bandgap (confirmed by a series of numerical simulations in COMSOL), where both modes experience significant losses and manifest near-zero effective permittivity, revealing the presence of a bandgap where plasmons propagation is suppressed. We emphasize that, because the domain of higher effective index modes in asymmetric waveguides has not previously been systematically analyzed, the phenomenon of this bandgap opening in non-periodic asymmetric 3L waveguides has not yet been observed. As an application example of the discovered bandgap, we discuss a novel approach to control the coupling between free-space excitation waves and surface plasmons via tapered non-adiabatic IMI waveguides.

## 2. Theoretical formalism

We first assume an electromagnetic wave propagating along the metal-insulator interfaces in a 3L geometry (plasmonic waveguide) in the form of $E(z)e^{i\beta x}$, using the field components of the wave equation for the TM modes $E_x = -i\frac{1}{\omega\varepsilon_0\varepsilon}\frac{\partial H_y}{\partial z}$, $E_z = -\frac{k_p}{\omega\varepsilon_0\varepsilon}H_y$, and applying boundary conditions for $H_y$ and $E_x$ at the interfaces, we derive a compact plasmonic dispersion relation for the general case of the asymmetric 3L plasmonic waveguide:

$$\tanh 2k_2 t + \frac{(k_1\varepsilon_3 + k_3\varepsilon_1)k_2\varepsilon_2}{k_1 k_3 \varepsilon_2^2 + k_2^2 \varepsilon_1 \varepsilon_3} = 0 \tag{1}$$

where $k_j^2 = k_p^2 - k_0^2 \varepsilon_j$, $k_p$ and $k_0$ are the wavevectors of a plasmon and an excitation wave in free space respectively, $\varepsilon_j$ is a dielectric permittivity of layer $j = 1,2,3$, and $\varepsilon_1 \neq \varepsilon_2 \neq \varepsilon_3$. In order to avoid any confusion and for the sake of simplicity, we select the indexation of layers to always be from top to bottom, assuming layers 1 and 3 to be semi-infinite. Layer 2 is a thin film of thickness $t$. The waveguide is assumed to be infinite in the $y$ direction, and all materials are isotropic and non-magnetic. Different forms of three-layer dispersion can be found, for example, in [7, 13, 20, 25, 30].

Eq. (1) can be rewritten as:

$$\tanh k_2 2t + \frac{\frac{k_2\varepsilon_3}{k_3\varepsilon_2} + \frac{k_2\varepsilon_1}{k_1\varepsilon_2}}{1 + \frac{k_2\varepsilon_3}{k_3\varepsilon_2}\frac{k_2\varepsilon_1}{k_1\varepsilon_2}} = 0 \tag{2}$$

Then, using $\tanh(x+y) = \frac{\tanh(x) + \tanh(y)}{1 + \tanh(x)\tanh(y)}$, Eq. (2) can be transformed into the following form:

$$\tanh 2k_2 t = \tanh\left(\operatorname{arctanh}\left(-\frac{k_2 \varepsilon_1}{k_1 \varepsilon_2}\right) + \operatorname{arctanh}\left(-\frac{k_2 \varepsilon_3}{k_3 \varepsilon_2}\right)\right) \tag{3}$$

$$\text{or } 2k_2 t = \operatorname{arctanh}\left(-\frac{k_2 \varepsilon_1}{k_1 \varepsilon_2}\right) + \operatorname{arctanh}\left(-\frac{k_2 \varepsilon_3}{k_3 \varepsilon_2}\right) + i\pi q \tag{4}$$

where $q$ is an integer. Applying $\operatorname{arctanh}(x) = \operatorname{arctanh}\left(\frac{1}{x}\right) + i\frac{\pi}{2}$ Eq. (4) can be rewritten as:

$$k_2 2t = \operatorname{arctanh}\left(-\frac{k_2 \varepsilon_1}{k_1 \varepsilon_2}\right) + \operatorname{arctanh}\left(-\frac{k_2 \varepsilon_3}{k_3 \varepsilon_2}\right) + 2\pi m i \tag{5}$$

$$k_2 2t = \operatorname{arctanh}\left(-\frac{k_1 \varepsilon_2}{k_2 \varepsilon_1}\right) + \operatorname{arctanh}\left(-\frac{k_3 \varepsilon_2}{k_2 \varepsilon_3}\right) + 2\pi l i \tag{6}$$

where $m$, $l$ are integers; the solutions only of $m = 0$ and $l = 0$ are the low-loss eigenmodes. In the case of the IMI waveguide, Eqs. (5) and (6) correspond to modes with even symmetry and odd symmetry, respectively. Similar dispersion relations were demonstrated in [30], but derived in an alternative way.

Specifically, when the thickness $t$ of the thin film in the 3L waveguide is quite small, we exclusively deal with the high effective index modes satisfying $k_0^2 \varepsilon_j / k_p^2 \sim 0.01$; thus, we can assume $k_j^2 \approx k_p^2$, and then simplify the dispersion Eqs. (1), (5), and (6) to the following compact and explicit form:

$$\tanh 2k_p t + \frac{(\varepsilon_1 + \varepsilon_3)\varepsilon_2}{\varepsilon_2^2 + \varepsilon_1 \varepsilon_3} = 0 \tag{7}$$

$$k_p 2t = \operatorname{arctanh}\left(-\frac{\varepsilon_1}{\varepsilon_2}\right) + \operatorname{arctanh}\left(-\frac{\varepsilon_3}{\varepsilon_2}\right) + 2\pi m i \tag{8}$$

$$k_p 2t = \operatorname{arctanh}\left(-\frac{\varepsilon_2}{\varepsilon_1}\right) + \operatorname{arctanh}\left(-\frac{\varepsilon_2}{\varepsilon_3}\right) + 2\pi l i \tag{9}$$

In contrast to optical waveguides, where guided modes are characterized by a phase and a group velocity because of the material dispersion, the material dispersion of the plasmonic waveguides does not play such a crucial role, even in distances of several hundreds of plasmonic oscillations [31, 32]. Thus, modes excited at some frequency $k_0$ can be considered as single frequency plasmon modes. Therefore, we adopt the following definitions: plasmons, exhibiting a positive effective index ($\operatorname{Re}\{k_p\}>0$, $\operatorname{Im}\{k_p\}>0$), have parallel energy flow and phase velocity, while plasmons with a negative index ($\operatorname{Re}\{k_p\}<0$, $\operatorname{Im}\{k_p\}>0$) have oppositely directed phase velocity and energy flow [7, 8, 20, 33].

### 3. Plasmonic bandgap phenomenon

As can be seen from Eqs. (8) and (9), the dispersion behavior for both the positive and negative index modes is critically altered when $|\operatorname{Re}\{\varepsilon_2\}|=|\operatorname{Re}\{\varepsilon_1\}|$ or $|\operatorname{Re}\{\varepsilon_2\}|=|\operatorname{Re}\{\varepsilon_3\}|$ ($|\varepsilon_2|=|\varepsilon_1|$ or $|\varepsilon_2|=|\varepsilon_3|$ for the lossless case), exhibiting a step-like increase of losses for the positive index mode (Fig. 1(a), Eq. (8)), or decrease of these for the negative index mode (Fig. 1(b), Eq. (9)). Specifically, when the *arctanh* arguments of Eqs. (8) and (9) become greater than 1, the frequencies (momenta) of $k_0$, in which both even and odd mode losses experience abrupt changes (will be denoted as $k_0^{|\varepsilon_1|=|\varepsilon_2|}$ and $k_0^{|\varepsilon_3|=|\varepsilon_2|}$), are calculated from the $|\varepsilon_2|=|\varepsilon_1|$ and $|\varepsilon_2|=|\varepsilon_3|$ conditions using a Drude-Lorentz model $\varepsilon_m = \varepsilon_b - \omega_p^2/(\omega^2 + \omega\gamma i)$:

$$k_0^{|\varepsilon_d|=|\varepsilon_m|} = \frac{1}{c}\sqrt{\frac{\omega_p^2}{\varepsilon_d + \varepsilon_b} - \gamma^2} \tag{10}$$

where $\varepsilon_d$ is a dielectric permittivity of the insulator layer $j = 1,3$ in the IMI ($j = 2$ in MIM) waveguide and $\varepsilon_m$ is a dielectric permittivity of the metallic layer $j = 2$ in the IMI ($j = 1,3$ in MIM) waveguide. $\varepsilon_b$, $\omega_p$, and $\gamma$ are the background permittivity, relaxation rate, and plasma frequency of the Drude-Lorentz model, respectively, for metals and heavily doped semiconductors.

For example, in the IMI geometry with $\varepsilon_1 < \varepsilon_3$, the plasmonic mode associated with the positive index has a low damping region at frequencies of $k_0 < k_0^{|\varepsilon_3|=|\varepsilon_2|}$ (Fig. 1(a)), while the mode associated with the negative index exhibits low losses at $k_0 > k_0^{|\varepsilon_1|=|\varepsilon_2|}$ (Fig. 1(b)). Between the low-loss spectral regions, a spectral zone ($k_0^{|\varepsilon_3|=|\varepsilon_2|} > k_0 > k_0^{|\varepsilon_1|=|\varepsilon_2|}$) occurs, where only highly damped solutions exist (blue shaded areas in Fig. 1(c)). In terms of a plasmon energy transfer, high losses of these modes can be related to equal and oppositely directed energy flows in the insulator and metal layers, which is reflected in the *near*-zero values of the effective permittivity for both modes $\varepsilon^{eff}(k_0) = \int |E_x(z)|^2 \varepsilon(k_0,z)dz / \int |E_x(z)|^2 dz$, where $|E_x(z)|^2$ is the intensity distribution of the electrical field of the mode (Fig. 1(d)). The large imaginary part of the mode wavevectors in this spectral region, along with near-zero $\varepsilon^{eff}$, indicate the existence of a bandgap where the plasmons propagation is suppressed. The low-loss merged dispersion profile for the eigenmodes in the IMI structure with the bandgap is shown in Fig. 1(c). A similar bandgap phenomenon can be observed in the asymmetric MIM waveguides. It should be noted that despite the seeming similarity between the $\bar{n}$-zero bandgap in photonic crystals [4, 5] and currently demonstrated plasmonic bandgap with a $\varepsilon$-zero property, the behaviors of the guided modes in each bandgap differ. The main difference can be observed through mode wavevectors that possess both finite real and imaginary components within the plasmonic bandgap (i.e. plasmon propagation is suppressed because of high losses), while in the *n*-zero bandgap, the wavevector has only an imaginary part [4, 5] (i.e. light wave propagation is impossible since it experiences zero group velocity).

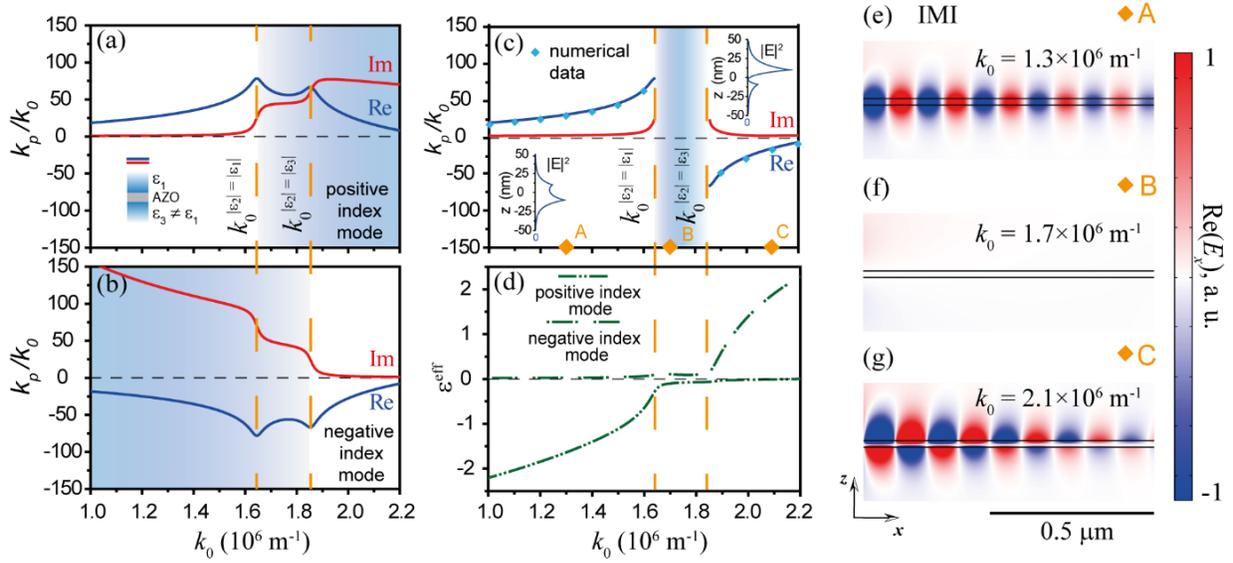

Fig. 1. Dispersion of the IMI plasmonic modes associated with positive (a) and negative (b) effective indices. (c) Merged dispersion, where the blue symbols illustrate the results of a numerical simulation; (insets) typical field distribution of each mode (positive mode at the left bottom, negative mode at the right top). (d) Effective value of permittivity for positive and negative plasmonic modes, in an asymmetric IMI structure. (e)-(f) Numerically obtained snapshots of the electric field distribution ($E_x$), associated with the plasmon modes of positive index (e), in the bandgap (f), and negative index (g). The dielectric permittivities of layers in all cases are $\varepsilon_1 = 2.25$ and $\varepsilon_3 = 3.9$; the thickness of the AZO layer is assumed to be 20nm.

A series of numerical simulations was conducted using COMSOL finite element method (FEM) software in order to confirm the existence of the bandgap, and the results are shown as blue data points in Fig. 1(c). The dielectric permittivities of the insulator layers are assumed to be $\varepsilon_1 = 2.25$ and $\varepsilon_3 = 3.9$ (BK7 and SiO$_2$), while the heavily Al-doped zinc oxide (AZO) semiconductor was chosen as the metallic material. We consider the following physical properties of the AZO layer: doped carrier concentration $n = 2\times10^{20}$ cm$^{-3}$, $\varepsilon_b = 3.8$, charge carriers mobility $\mu = 60$ cm$^2$/Vs, and electron effective mass $m^* = 0.34\times m_e$ [34-39].

Electric field distributions of plasmonic modes along the $x$-axis in the IMI waveguide for three excitation wavelengths (marked as orange data points A, B, and C on the $k_0$ axis in Fig. 1(c)) are demonstrated in Figs. 1(e)-1(g). Each of these points A, B or C is referred to a specific dispersion window (Fig. 1(c)), demonstrating a typical plasmonic response thereof. For example, point A ($\lambda_0 = 4.83$ μm) demonstrates a positive effective index plasmonic mode of even symmetry (Fig. 1(e)), point B ($\lambda_0 = 3.7$ μm) is located in the bandgap zone, showing the absence of excitations (Fig. 1(f)), while point C ($\lambda_0 = 3$ μm) is associated with a negative effective index mode of odd symmetry (Fig. 1(g)). Typical electric field intensity profiles of these modes are shown in the inset of Fig. 1(c).

## 4. Plasmonic bandgap properties

An important feature of the plasmonic bandgap in non-periodic asymmetric waveguides is the capability of control over its spectral width (Fig. 2(a)) and position (Fig. 2(b)) by an appropriate choice of employed materials. In the IMI waveguide, the bandgap's width can be controlled through the dielectric permittivity of insulator layers, while its spectral position can be determined solely with the selection of the metallic layer material (see Eq. (5)). Fig. 2(b) demonstrates spectral lines $k_0^{|\varepsilon_d|=|\varepsilon_m|}$ satisfying $|\varepsilon_m|=|\varepsilon_d|$ in the 3L waveguide for the most popular plasmonic materials: metals, metalic graphene, and some semiconductors. A spectral gap between any two points $k_0^{|\varepsilon_1|=|\varepsilon_2|}$ and $k_0^{|\varepsilon_3|=|\varepsilon_2|}$ on the spectral line (lines) presents the plasmonic bandgap window in the 3L waveguide. As can be seen in Fig. 2, the width and position of the spectral gap can be tuned in a spectral range from infrared to ultraviolet. This shows that our approach of selective bandgap tuning represents a powerful instrument for the design of IMI/MIM plasmonic waveguides with desired properties across the broad spectral range.

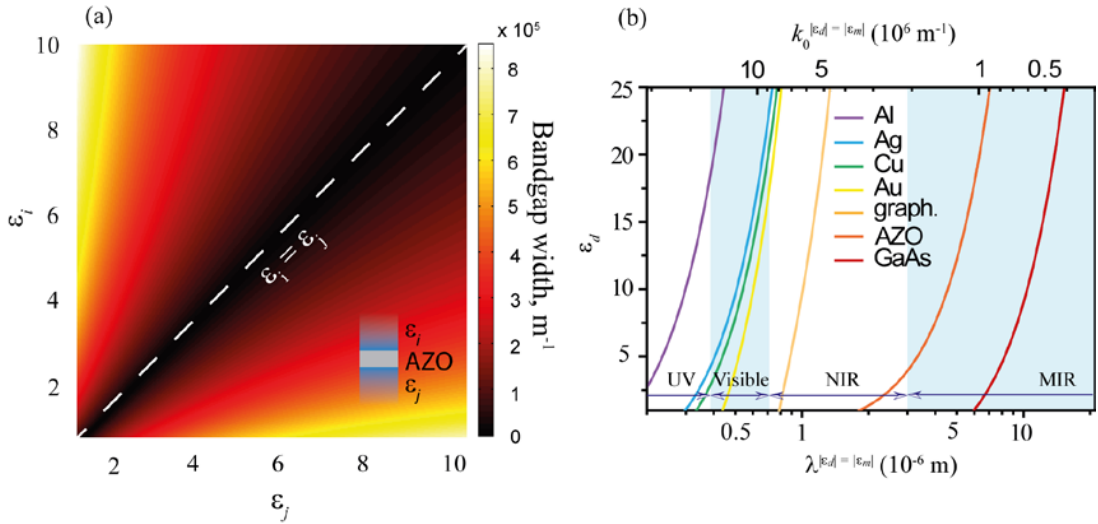

Fig. 2 (a) Spectral width (m$^{-1}$) of the dispersion bandgap as a function of dielectric permittivities of top and bottom layers $\varepsilon_i$ and $\varepsilon_j$ in the IMI geometry, with an example of AZO film in the middle layer. The line $\varepsilon_i = \varepsilon_j$ corresponds to zero bandgap width. (b) Momentum $k^{|\varepsilon_d|=|\varepsilon_m|}$ for different dielectric permittivities $\varepsilon_d$ and plasmonic materials. Metal-like graphene is considered to be doped up to 1 eV, AZO having a charge carriers concentration of $2\times10^{20}$ cm$^{-3}$ and GaAs of $10^{19}$ cm$^{-3}$.

## 5. Perspective applications

A new potential application of the presented bandgap phenomenon can be found in the tapered non-adiabatic IMI waveguide [30, 40] for its significant reflections and scattering can be eliminated with the introduction of a bandgap near the tip of the waveguide. A plasmon, excited on the *thick* part of such waveguide, is a regular surface plasmon at the IM interface (excited at frequency $k_0$ and having wavenumber $k_p$, see Fig. 3(a)). As it approaches the tapered end of the waveguide, in commonly considered symmetric (or weakly asymmetric) tapered waveguides, this mode is transformed into a high effective index mode of the IMI structure (i.e. nanofocusing) [13,30,40,41], experiencing reflection (especially in non-adiabatic waveguides) which produces an interference pattern of the energy density $W$ (Fig 3(b)) [40,42,43]. In sharp contrast, for the case of strongly asymmetric waveguides with $\varepsilon_3 > \varepsilon_1$, a phenomenon of unidirectional radiation in the surrounding medium can be observed (Fig. 3(c)). This phenomenon is related to the coupling of the single interface mode in a leaky long range surface plasmon mode (LRSP) which is further coupled in a radiative mode (see appendix). The angle between the wavevectors of the plasmonic and radiative modes at the tip of the waveguide can be calculated as $\cos\beta = \sqrt{\varepsilon_1}/\sqrt{\varepsilon_3}$ (see appendix). The coupling in the radiative mode is considerably enhanced (up to two orders of magnitude) in the bandgap zone due to the suppression of the 3L waveguide modes and thus the reflection at the tip (Fig. 3(d)).

Similarly, the mechanism of light coupling into surface plasmons at a single interface can be realized (Fig. 3(h)) with suggested strongly asymmetric waveguides, providing a new technique of plasmons excitation along with well-known tools such as prisms, gratings, etc. [17, 20]. Numerical simulation results reveal the efficient coupling of a selective excitation wave in surface plasmons (Fig. 3(g)) and the LRSP, achieving an efficiency of up to 20% in non-adiabatic IMI waveguides not relying on complex geometries or multiple radiation sources [18, 19, 44].

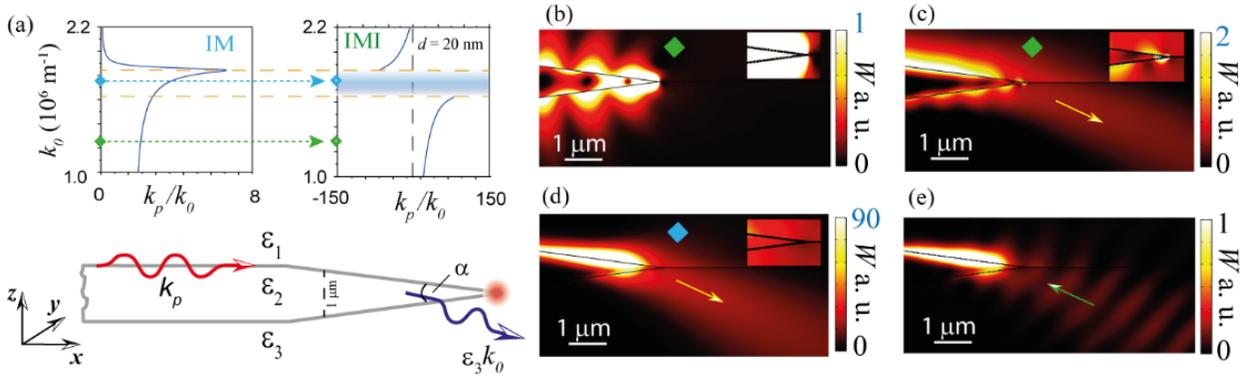

Fig. 3 (a) Schematics of light emission mechanism using a tapered IMI waveguide with tapered angle $\alpha$ = 15 degrees considered as non-adiabatic (used in all simulations) [30, 40]. (b) Nanofocusing for $k_0 = 1.25\times10^6$ m$^{-1}$ in symmetric waveguide with $\varepsilon_1 = \varepsilon_3 = 2.25$. (c) Unidirectional light emission along with nanofocusing for $k_0 = 1.25\times10^6$ m$^{-1}$ in asymmetric waveguide with $\varepsilon_1 = 2.25$ and $\varepsilon_3 = 3.9$. (d) Similar to (c), except for frequency (in the bandgap, $k_0 = 1.75\times10^6$ m$^{-1}$). (e) Light coupling in the tapered end of the IMI waveguide (for $k_0 = 1.75\times10^6$ m$^{-1}$, at the direction of maximum coupling intensity).

## 6. Conclusion

We demonstrate that the plasmonic dispersion of an asymmetric IMI waveguide is characterized by step-like changes of the wavevector's imaginary part, for their two co-existing modes. This leads to the formation of the spectral bandgap, where plasmons propagation is suppressed because of the high losses and near-zero effective permittivity for supported modes. The bandgap width and position can be readily adjusted by selecting employed materials, allowing the excitation of plasmons with positive and negative effective indices at a very broad spectrum range from UV to IR. The plasmonic bandgap phenomenon has bright perspectives as a new technique of effective surface plasmon excitation and light emission on tapered waveguides. Our results provide an important contribution to the field of plasmonic waveguide design and applications, such as detectors fabrication, antennas, near-field microscopy, and nanosensing.

## 7. Appendix

Fig. 4(a) shows the plasmonic dispersion curves of the single IM interface (mode "0", in green) and IMI waveguide (modes "1","2" and "3"), where mode "1" (orange) is the positive effective index mode, mode "2" (blue) is the negative effective index mode, and mode "3" (pink) is the long range surface plasmon (LRSP) mode [24].

Generally, in a tapered IMI waveguide, the IM interface mode, starting from the thick part of the waveguide in the direction of the tip, is continuously coupled in the IMI waveguide modes, while the thickness $t$ approaches zero. However, in the bandgap spectral zone, such mode transition is favorable only between the single IM interface ($\varepsilon_1$|AZO) mode "0" and the LRSP mode "3", since other modes "1" and "2" become extremely lossy. Fig. 4(b) demonstrates the smooth transition of the dispersion solution for the single IM interface mode (point A) towards the LRSP mode (point B) as a function of the waveguide's middle layer thickness $t$, at fixed frequency $k_0 = 1.8 \times 10^6$ m$^{-1}$. The LRSP mode is characterized by a cutoff momentum $k^{cutoff}$ and a cutoff thickness $t^{cutoff}$ [24, 27]. We show in Figs. 4(a) and 4(b) that $k^{cutoff}$ as well as $t^{cutoff}$ are only transition points of the LRSP mode "3" in the leaky regime. The leaky LRSP mode ($k_0 < k^{cutoff}$; $t < t^{cutoff}$) exists between two light lines $k_0\sqrt{\varepsilon_1}$ and $k_0\sqrt{\varepsilon_3}$ (asymptotically approaching $k_0\sqrt{\varepsilon_1}$, when $\varepsilon_1 < \varepsilon_3$), and thus can be directly excited by an incident light wave from one of the dielectric layers in the IMI waveguide.

The efficient coupling between the low-loss modes supported by the tapered waveguide (IM interface – LRSP – Leaky LRSP – radiative) is a basis mechanism behind the demonstrated free space emission/excitation of plasmons.

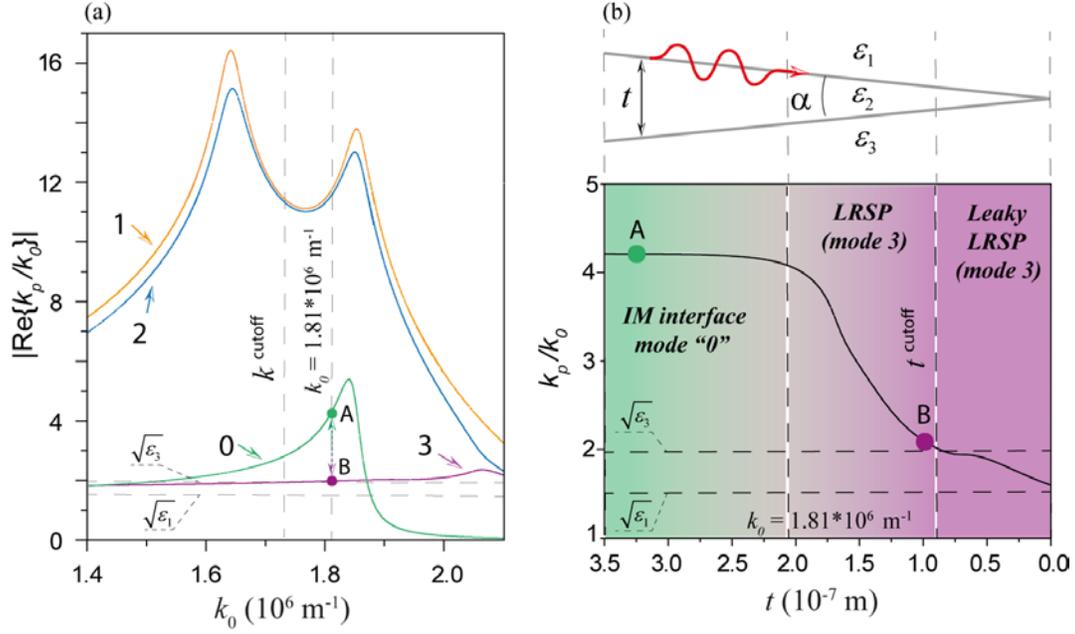

Fig 4. (a) Plasmonic response (*absolute* value of Re{$k_p/k_0$}) of IM interface with $\varepsilon_1=2.25$ and AZO (mode 0 of IM interface), and IMI geometry with $\varepsilon_1 = 2.25$, $\varepsilon_3 = 3.9$ and AZO film of thickness $t = 100$ nm (modes 1, 2 and 3). The horizontal dashed lines $k_p/k_0 = \sqrt{\varepsilon_1}$ and $k_p/k_0 = \sqrt{\varepsilon_3}$ are light lines of corresponding dielectrics. (b) Mode effective index $k_p/k_0$ as a function of film thickness $t$ at frequency $k_0 = 1.81\times10^6$ m$^{-1}$ demonstrating a smooth transition of the single IM interface plasmonic mode in the (leaky) LRSP mode. The light lines shown are the same as in (a).

Due to the asymptotic behavior of LRSP (when the thickness of the waveguide decreases), $k_p/k_0$ approaches the light line value $\sqrt{\varepsilon_1}$ (when $\varepsilon_1 < \varepsilon_3$), while the radiative mode propagates into the media with $\sqrt{\varepsilon_3}$ (when $\varepsilon_1 < \varepsilon_3$), making it possible to calculate the angle between the wavevectors of these two modes at the tip of the waveguide:

$$cos\beta = \sqrt{\varepsilon_1}/\sqrt{\varepsilon_3} \quad (\varepsilon_1 < \varepsilon_3) \qquad (11)$$

In fact, the LRSP mode becomes leaky after $t^{cutoff}$, and the most effective radiation will consequently occur before reaching the tip and will be characterized by a slightly smaller effective $\beta^{eff}$ than that given by Eq. (11).

**Acknowledgments**

We thank Professor Peter Pikhitsa for valuable discussions. This work was supported by the National Research Foundation under the Ministry of Science, ICT & Future Planning, the Global Frontier Program NRF-2014M3A6B3063708, and the Global Research Laboratory (GRL) Program K20815000003 (2008-00580); all funded by the South Korean government.